\documentclass[aps,prl,nofootinbib,preprintnumbers,amsmath,amssymb,latexsym,array,enumerate,letter,twocolumn,superscriptaddress]{revtex4}
\usepackage{amssymb}
\usepackage{amsmath}
\usepackage{epsfig}
\usepackage{hyperref}
\usepackage{breakurl}
\usepackage{xcolor}
\usepackage{subfigure}
\usepackage{caption}
\usepackage{soul}
\usepackage{graphicx}

\makeatletter
\def\simgt{\mathrel{\lower2.5pt\vbox{\lineskip=0pt\baselineskip=0pt
           \hbox{$>$}\hbox{$\sim$}}}}
\def\simlt{\mathrel{\lower2.5pt\vbox{\lineskip=0pt\baselineskip=0pt
           \hbox{$<$}\hbox{$\sim$}}}}
\makeatother

\newcommand{\be}{\begin{equation}}
\newcommand{\ee}{\end{equation}}
\newcommand{\bea}{\begin{eqnarray}}
\newcommand{\eea}{\end{eqnarray}}
\newcommand{\beq}{\begin{eqnarray}}
\newcommand{\eeq}{\end{eqnarray}}

\def\lsim{\mathrel{\rlap{\lower4pt\hbox{\hskip1pt$\sim$}}
     \raise1pt\hbox{$<$}}}         
\def\gsim{\mathrel{\rlap{\lower4pt\hbox{\hskip1pt$\sim$}}
     \raise1pt\hbox{$>$}}}         

\begin{document}


\title{Producing and detecting long-lived particles at different experiments at the LHC}
\pagestyle{plain}
\author{Chaochen Yuan}
\affiliation{Institute of High Energy Physics, Chinese Academy of Sciences, Beijing, 100049, China}
\affiliation{University of Chinese Academy of Sciences, Beijing, 100049, China}

\author{Guoming Chen}
\affiliation{Institute of High Energy Physics, Chinese Academy of Sciences, Beijing, 100049, China}
\affiliation{University of Chinese Academy of Sciences, Beijing, 100049, China}

\author{Huaqiao Zhang\footnote{Corresponding author: zhanghq@ihep.ac.cn}}
\affiliation{Institute of High Energy Physics, Chinese Academy of Sciences, Beijing, 100049, China}
\author{Yue Zhao\footnote{Corresponding author: zhaoyue.hep@gmail.com}}
\affiliation{Department of Physics and Astronomy, University of Utah, Salt Lake City, UT 84112, USA }

\begin{abstract}
We propose a new strategy to look for long-lived particles (LLP) at the LHC. The LLPs are produced at one experiment, but its decay products are detected by a detector at another experiment. We use a confining Hidden Valley scenario as a benchmark. Through showering and hadronization, the multiplicity of hidden mesons can be large, and their decay products, dimuon as chosen in this study, are typically too soft to pass triggers in traditional LHC searches. We find the best acceptance is achieved if we produce LLPs at collision points at the LHCb and ALICE experiments, and use the muon chamber of ATLAS for detection. This new search is cost-efficient since it does not require a new detector to be built. Meanwhile, it can provide coverage of interesting parameter space, which is complementary to other proposed LLP searches.
\end{abstract}

\maketitle

\section{Introduction}

The LHC is the most energetic machine at the high energy frontier. There have been many measurements conducted at the LHC, however, most of them are specialized in looking for new particles that are produced and decay promptly in the same detector. By contrast, the existence of long-lived particles commonly appears as a natural prediction in many well-motivated frameworks of new physics beyond the SM \cite{Giudice:1998bp,Barbier:2004ez,Chacko:2005pe,Burdman:2006tz,Meade:2010ji,Arvanitaki:2012ps,ArkaniHamed:2012gw,Craig:2015pha}. The searches for such long-lived particles (LLPs) is a very interesting and important research direction. However, such searches are quite challenging at the LHC because these events involving LLPs are generically difficult to trigger, also the modeling of the relevant SM background is highly non-trivial.

Recently, much effort has been devoted to extending the LHC experiment by new detectors, such as SHiP \cite{Alekhin:2015byh}, FASER \cite{Feng:2017uoz}, MATHUSELA \cite{Chou:2016lxi}, 
and CODEX-b \cite{Gligorov:2017nwh}. All these proposals require adding additional facilities beyond the ones we have already built, which can be expensive and time consuming. While it is very important to push such extensions forward, it is interesting to explore the possibilities to probe part of the unexplored parameter space of LLP using new search strategies with the existing detectors at the LHC, of course not necessarily to have a comparable sensitivity as that can be reached by a delegated experiment.

\begin{figure}[htbp]
 \centering
 \includegraphics[width = .4\textwidth]{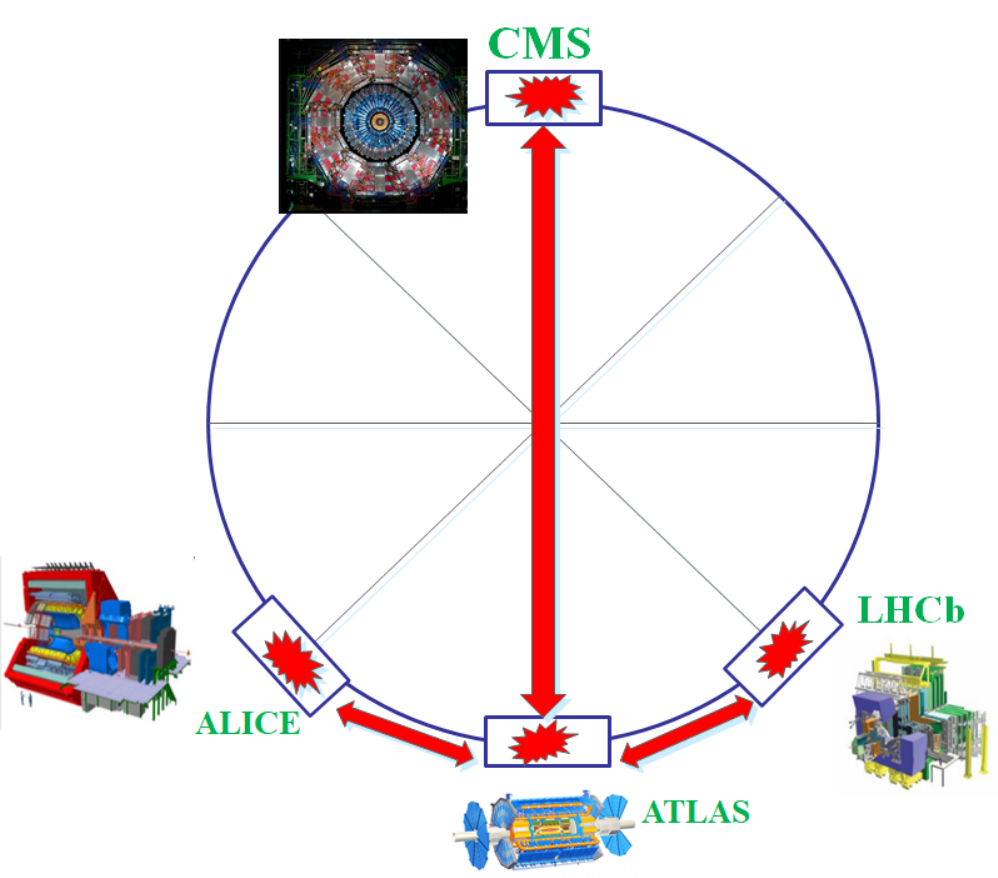}
 \caption{This is a bird-view of the LHC. Long-lived particles can be produced at one experiment and then detected at another experiment far away. }
 \label{LHC-diagram}
 \end{figure}

At the LHC, there are several independent experiments operating simultaneously, their detectors are specifically designed with different goals. In most studies carried out at the LHC, the detector is used to study the collisions happening at the collision point within itself. However, if LLPs exist and are produced by high energy collisions, they can escape the detector at the collision point and accidentally decay in the vicinity of detectors far away. In this case, the decay products, such as a dimuon pair, can leave largely displaced tracks in the far detector as they enter in, and then can potentially be used to reconstruct the long-lived mother particle.  For example, a high energy collision at the LHCb or ALICE experiment may produce multiple LLPs. One of them can decay to a dimuon pair in the vicinity of ATLAS, and these muons will leave their tracks in the ATLAS muon chamber, as illustrated in Fig. \ref{LHC-diagram}. 

To record LLP signal events, the far detector needs to trigger on charged tracks propagating from outside to inside, with large impact parameter. Special triggering strategy using Kalman Filter method has been tested and demonstrated to be working using current firmware of CMS detector~\cite{Foudas:2019uas}. This trigger strategy could be further optimized for High-Luminosity LHC data taking, based on the existing upgrade plan.





In this letter, we study details of this novel LLP search strategy, i.e. looking for LLPs produced from collision points of other experiments. This search will provide significant coverage of unexplored parameter space, complementary to that covered in existing LLP searches at the LHC. Furthermore, this new search strategy does not require additional hardware or facility being built, which is cost-efficient.


\section{Theory Setup}
LLPs naturally appear in many natural extensions of the SM. One typical class of models with LLPs in the particle spectrum is the hidden valley (HV) scenario \cite{Strassler:2006im}. In this section, we provide details about one benchmark model in this scenario. We note that our proposed search strategy can be easily applied to study many other models beyond the benchmark model we choose here.

We consider a heavy gauge boson, $Z'$, of the extended group $U(1)'$  to be the portal connecting our SM particles with particles in HV sector. We assume both SM quarks and hidden quarks are charged under $U(1)'$. The Lagrangian can be written as
\begin{eqnarray}\label{Lag}
L\supset g_{Z'} Z'_\mu (Q_{SM}\bar q \gamma^\mu q+Q_{HV} \bar q_{h} \gamma^\mu q_h)+\frac{1}{2} m_{Z'}^2 Z'^2
\end{eqnarray}
Here we take the $U(1)'$ gauge coupling as $g_{Z'}$. The SM and HV quarks have charges $Q_{SM}$ and $Q_{HV}$ respectively. If $Q_{SM}\ll Q_{HV}$, $Z'$ dominantly decays to HV sector and the dijet constraints on $Z'$ at the LHC is largely relaxed \cite{Aaboud:2018fzt}. The other possible constraint is from monojet search \cite{ATLAS:2017dnw} since $Z'$ mainly decays to hidden sector particles which escape the detector. However this only imposes very mild constraint on the coupling between $Z'$ and SM particles, i.e. $g_{Z'}Q_{SM}\sim 0.2$ if $Z'$ is 200 GeV. This indicates an upper limit on $Z'$ production cross section at 14 TeV LHC as more than 1000 $pb$. We will see that the sensitivity that can be achieved using our proposed search strategy does a much better job.

We assume the HV sector is governed by a confining non-abelian gauge group, similar to SM QCD. HV quarks from $Z'$ decay experience showering and hadronization (SH) under the hidden gauge group. This distributes their energy to softer hidden mesons. The meson spectrum in HV sector depends on details of the non-abelian gauge group, such as the numbers of color, flavor, confinement scale and HV quark masses. For simplicity and concreteness, we assume two species of hadrons in low energy spectrum, $\omega_h$ and $\eta_h$, which are vector and pseudo scalar mesons with similar masses. We expect such spectrum to appear when there is no chiral symmetry breaking induced by confinement. By a simple counting of the number of degrees of freedom, we expect $N_{\omega_h}/N_{\eta_h}\sim 3$. These mesons can be long-lived and escape the detector at the production point.
 \begin{figure}[htbp]
 \centering
 \includegraphics[width = .4\textwidth]{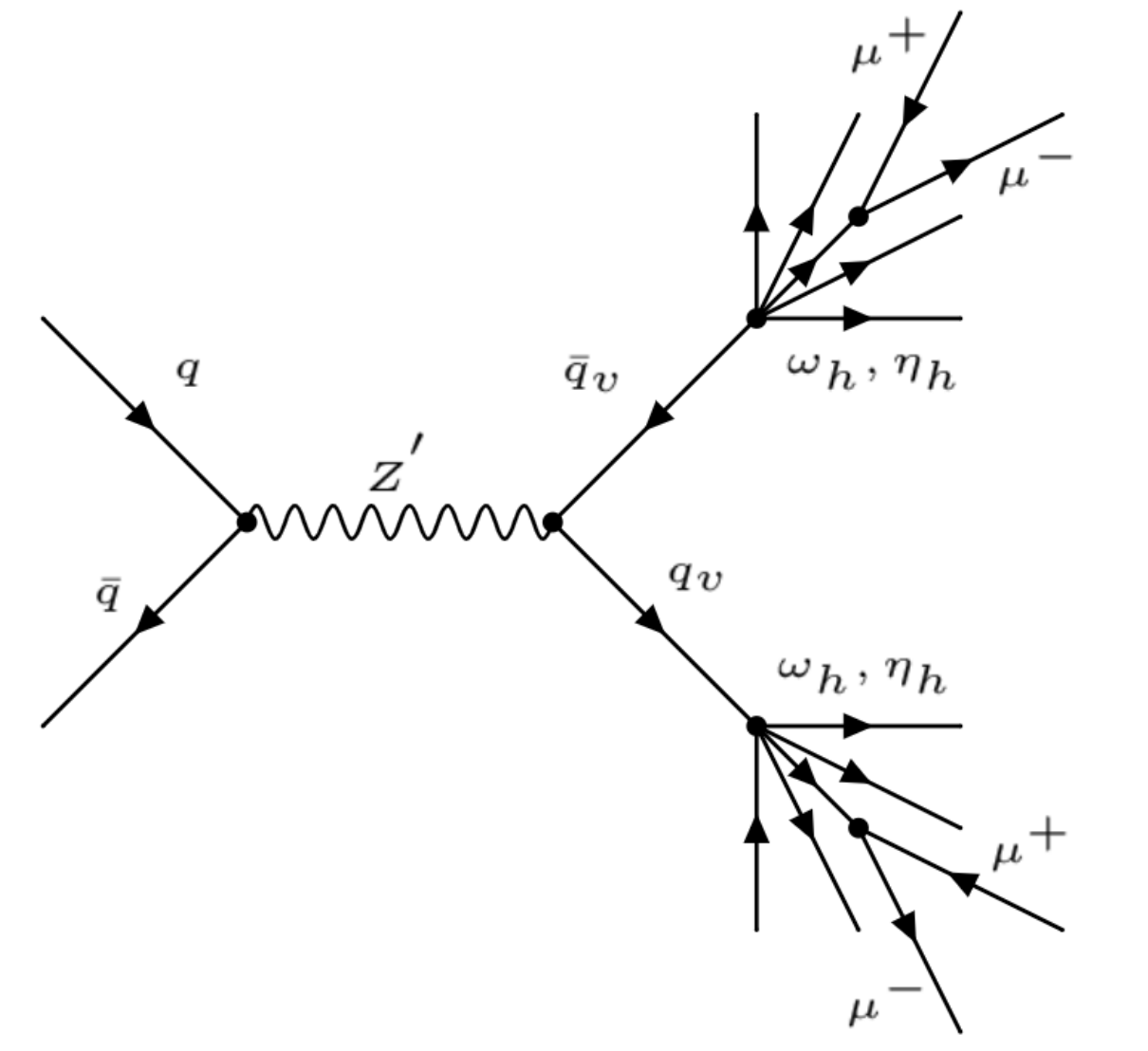}
 \caption{Leading order Feynman diagram for the process we are interested in}
 \label{Feynman diagram}
 \end{figure}

SH processes in hidden sector are highly non-trivial to model. Significant efforts have been devoted to simulate SH in HV scenario. As studied in \cite{Pierce:2017taw}, given a fixed number of HV mesons in final states, different SH techniques give approximately the same kinematics. For example, Pythia simulation \cite{Sjostrand:2014zea} gives a similar momentum distribution as that obtained using quark-combination model \cite{Xie:1988wi,Wang:1996jy,Si:1997rp} and the longitudinal phase space approximation \cite{Webber:1994zd,Han:2007ae}. Thus we adopt the simulation results from Pythia 8.1 in this study and treat the number of mesons as a free parameter.

After the production of these hidden mesons, we assume they decay back to SM sector eventually. The decay products and decay lifetime are also model dependent. Among all possible decay products, muon stands out for obvious reasons. First, muons can be well identified in the muon chamber, which is generally the largest component of a detector. This helps us to define our signal concretely. Furthermore, energetic muons can propagate for a long distance in earth. Thus the decays of HV mesons may happen on the way to a far detector and be registered by the far detector after a distant propagation. This largely improves the acceptance of our signal. In this study, we focus on the muon decay channel and our results scale linearly with the decay branching ratio to muons.

The decay lifetimes of $\omega_h$ and $\eta_h$ can be drastically different, depending on the type and coupling strength of mediating particles. In principle, one can tune parameters so that these two mesons share similar decay lifetimes, however, this is not generically the case. Here we focus on the decay of $\omega_h$'s and assume $\eta_h$'s have a much longer lifetime and most of them are not relevant for our study. Our results can be easily rescaled if one is interested in other possibilities.

\begin{figure}[htbp]
     \subfigure[ $m_{\omega_h}=0.3$ GeV ]{
     \includegraphics[scale=0.4]{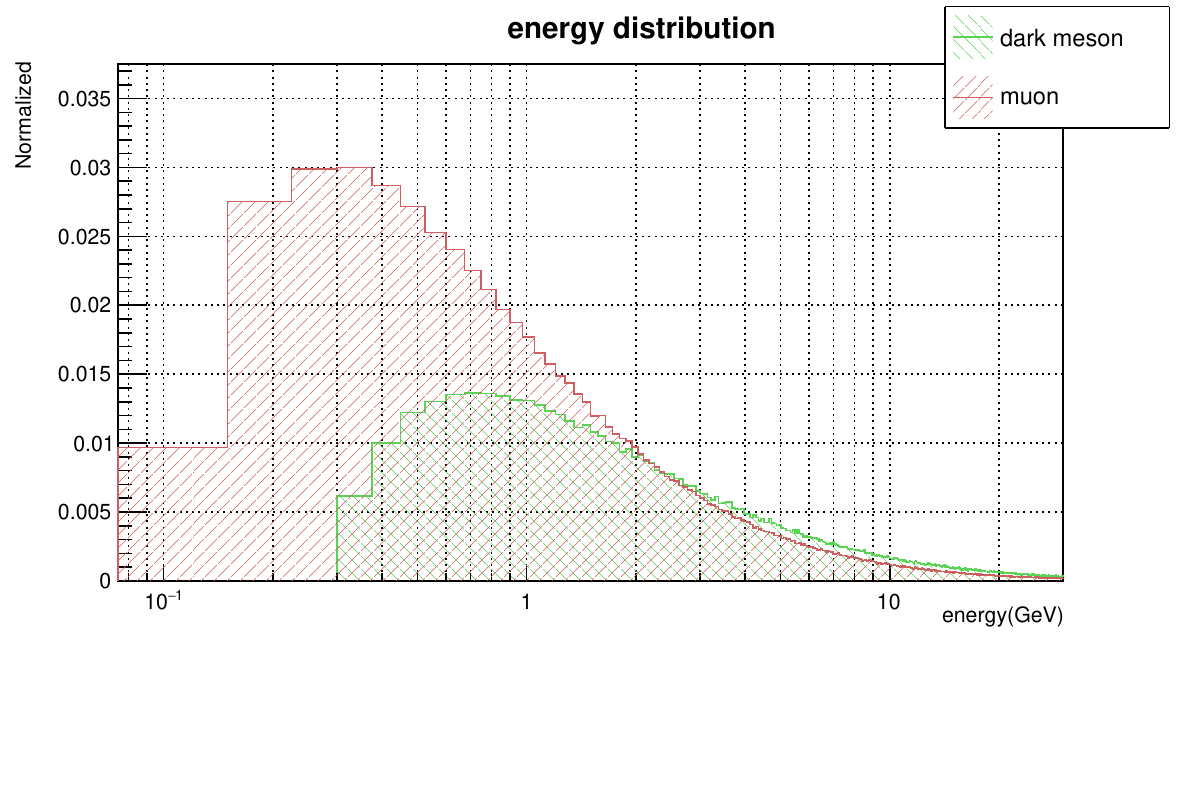}}
     \subfigure[ $m_{\omega_h}=5$ GeV]{
     \includegraphics[scale=0.4]{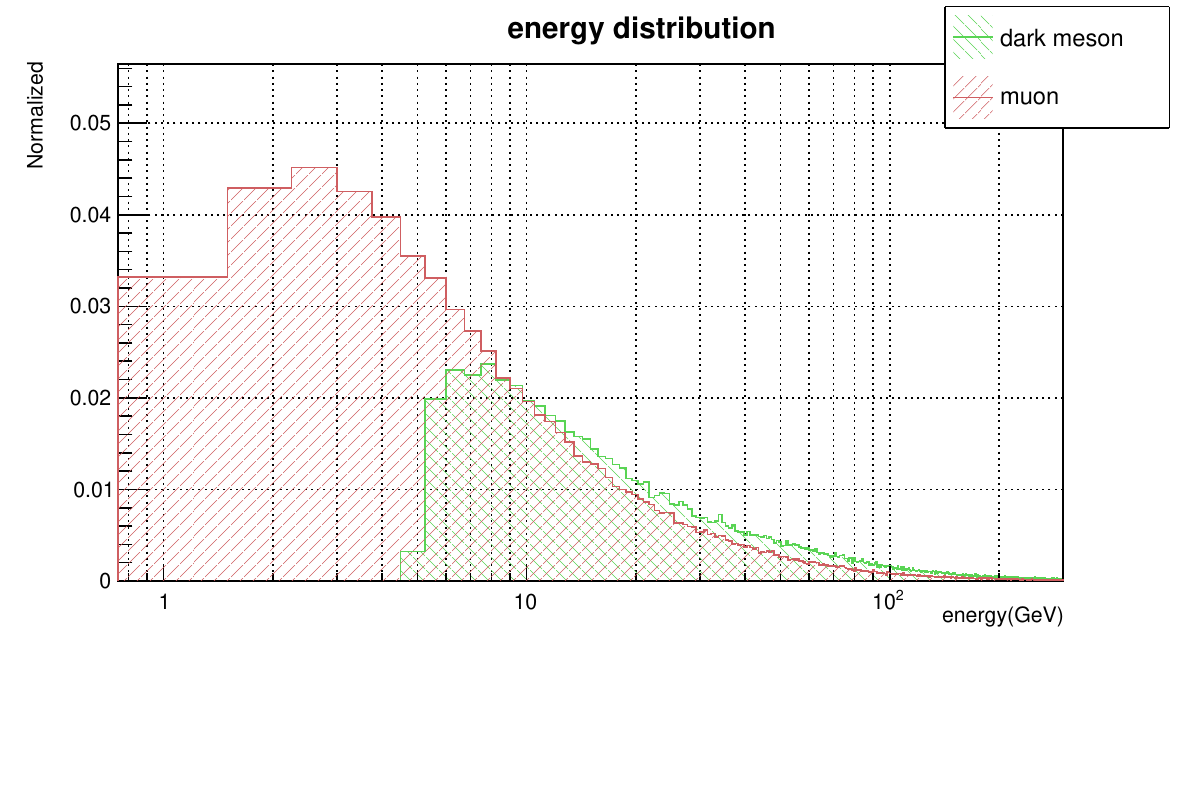}}
     \caption{Energy distributions of $\omega_h$ and muon with $m_{\omega_h}=0.3$ GeV and $m_{\omega_h}=5$ GeV respectively.  }
     \label{energy}
 \end{figure}

To gain some intuitions, we present several kinematic distributions of HV mesons and muons for a generic choice of benchmark parameters. The mass of $Z'$ is set to 200 GeV and the parameters in the hidden sector are picked to make the averaged HV meson number to be $\sim 40$ and $\sim 10$ when HV meson mass is 0.3 GeV and 5 GeV respectively. In Fig. \ref{energy}, we show the energy distributions of HV mesons and muons. 

Here we see that the boost factors of HV mesons are moderate. This indicates the directions of muons and the directions of HV mesons may not be perfectly aligned, especially when HV meson mass is much larger than muon mass, i.e. 5 GeV in our benchmark point. 

\begin{figure}[htbp] 
     \subfigure[$m_{\omega_h}=0.3$ GeV ]{
     \includegraphics[scale=0.4]{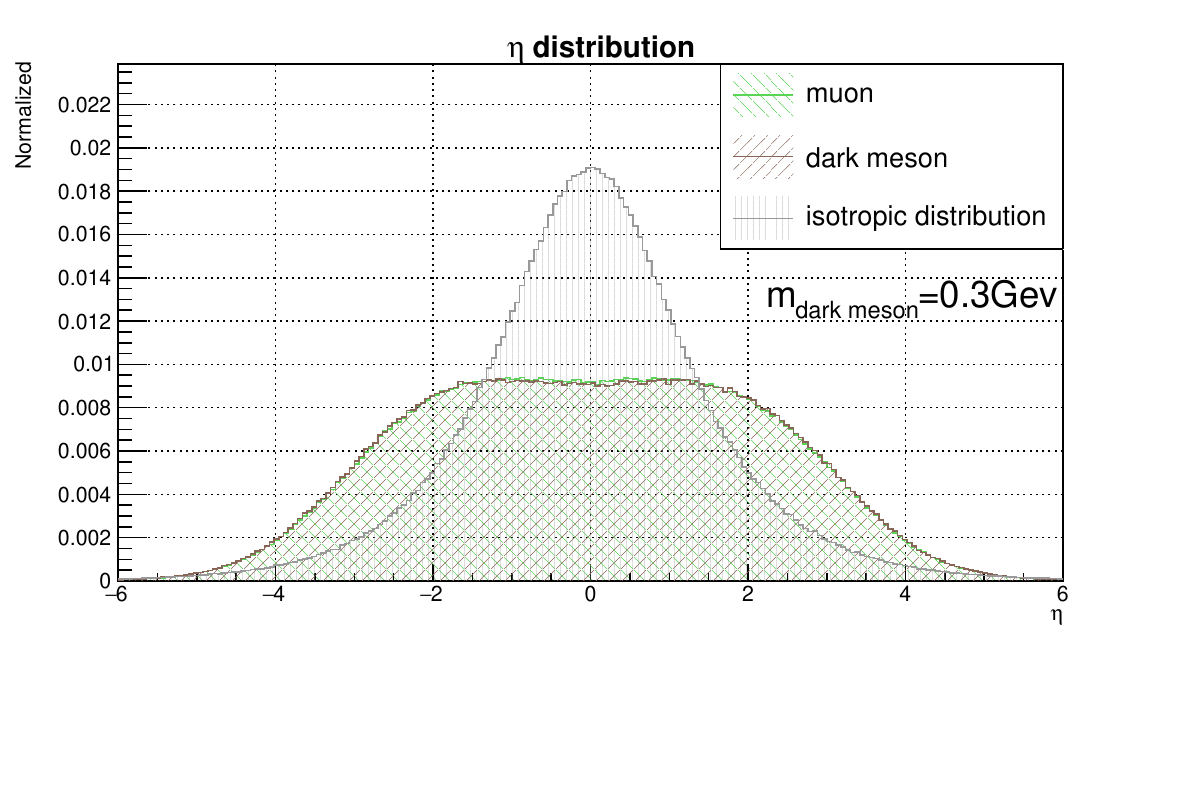}}
     \subfigure[$m_{\omega_h}=5$ GeV ]{
     \includegraphics[scale=0.4]{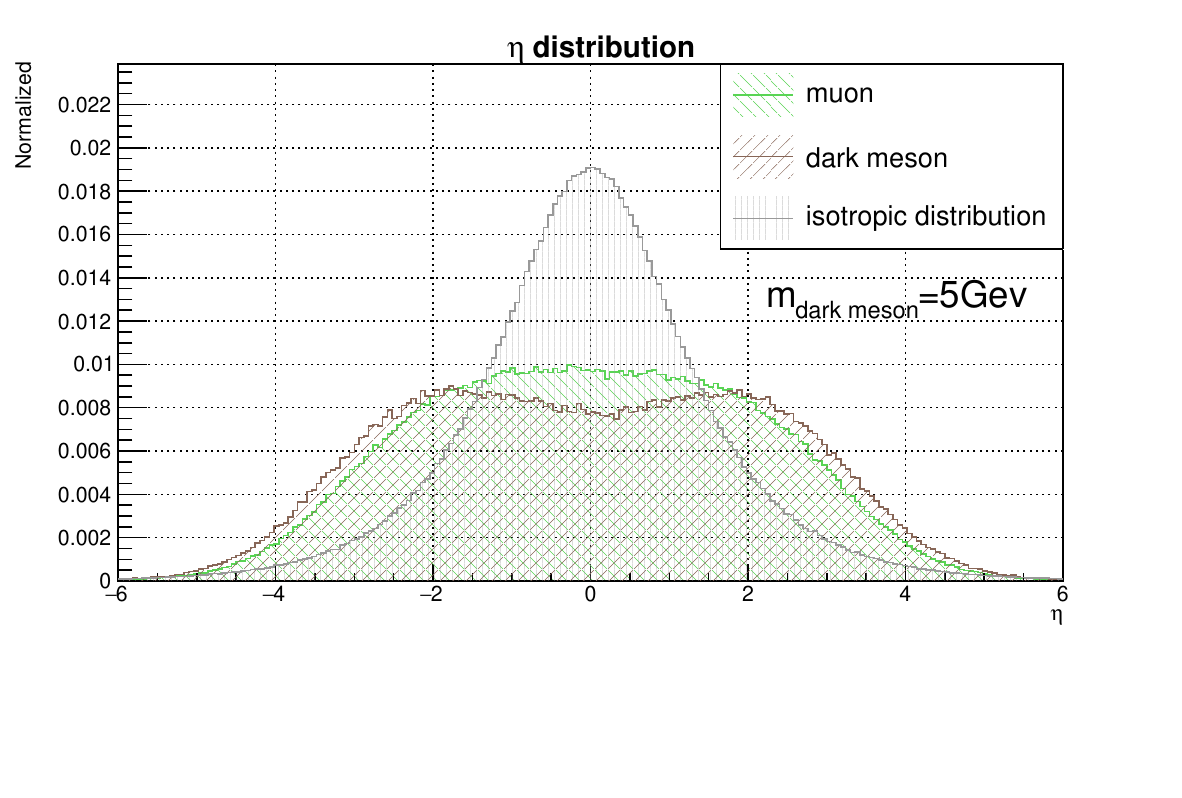}}
     \caption{$\eta$ distributions of $\omega_h$ and muon with $m_{\omega_h}=0.3$ GeV and $m_{\omega_h}=5$ GeV respectively. We also include an isotropic distribution as a reference.  }
     \label{angle}
 \end{figure}

In Fig. \ref{angle}, as a demonstration, we show the pseudo-rapidity ($\eta$) distribution of HV mesons and muons in the coordinates of the experiment at production. Here we also see that, rather than an isotropic distribution, the HV mesons are more concentrated along the beam direction, which is within expectation due to parton distribution function of a proton. This directional distribution indicates that a far detector located along the forward-backward direction can provide larger acceptance to our signal events.

\begin{figure}[htbp]
 \centering
 \includegraphics[width = .46\textwidth,height=.3\textwidth]{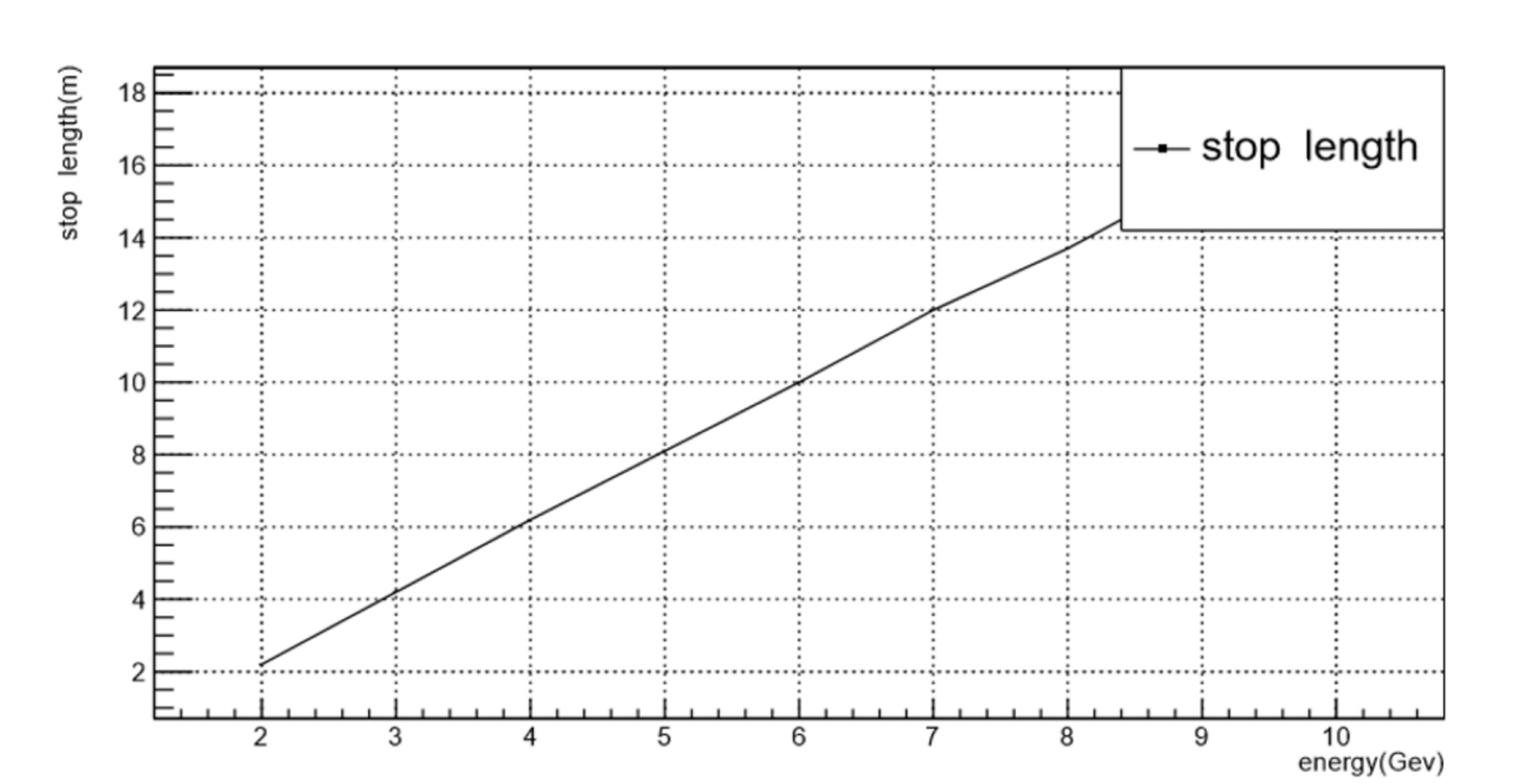}
 \caption{Average stop length as a function of muon energy.}
 \label{stoplength}
 \end{figure}

\begin{figure}[htbp]
 \centering
 \includegraphics[width = .5\textwidth,height=.35\textwidth]{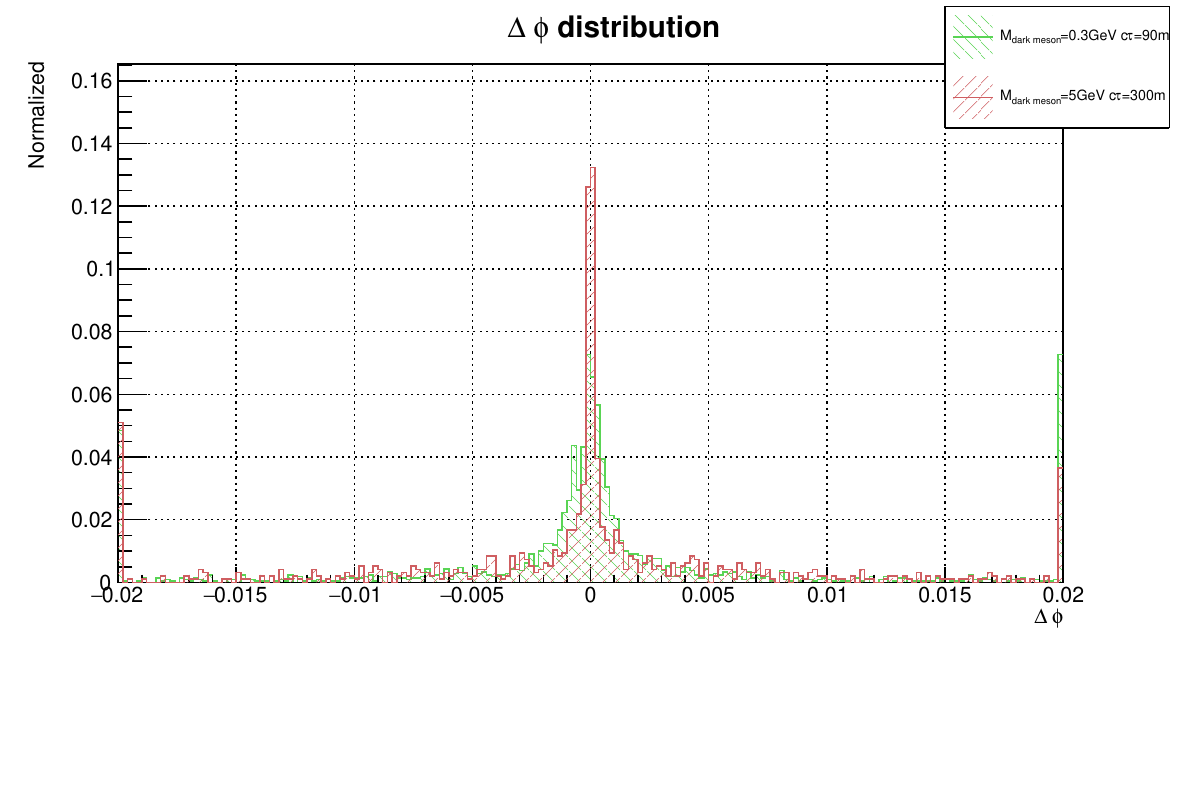}
 \caption{The change of muon directions after propagation in earth. }
 \label{delphi}
 \end{figure}

The propagation in earth of muons decayed from HV mesons is studied using MUSIC~\cite{Antonioli:1997qw,Kudryavtsev:1999zu}. In Fig. \ref{stoplength}, we show the averaged stop length of a muon as a function of its energy.  We further simulate the earth induced change on muon direction after they run into the chamber of the far detector, presented in Fig. \ref{delphi}. Here we require the muon to be energetic enough for detection ($p_T > 1$ GeV in the coordinates of the experiment at detection). As benchmarks, we choose to have $c\tau$ = 90 m for 0.3 GeV and 300 m for 5 GeV, where the best sensitivity can be achieved as shown later. $\Delta \phi$ is the change of the azimuthal angle for the muons in the coordinates of the production experiment. We note that the propagation in earth does not change the direction of a muon significantly. 

\section{Experimental setup}

In this study, we consider the detection method where the LLPs are produced at one LHC collision point (production site) and detected by an experiment at another collision point (detection site). 
There are 4 major experiments at the LHC. Each experiment has a detector sitting at one collision point of the LHC. The acceptance of LLP largely depends on the relative solid angle of a detector w.r.t. the collision point. 

The locations of the LHC experiments are shown in Fig. \ref{LHC-diagram}. The distance between each pair of experiments is listed in Table \ref{table1}.
\begin{table}[htbp!]
\centering
\begin{tabular}{|c|c|c|}
\hline
Produce &  Detect & Distance(m)\\
\hline
ALICE/LHCb & ATLAS & 3288\\
\hline
ALICE/LHCb & CMS & 7940\\
\hline
ATLAS & CMS & 8594\\
\hline
ALICE & LHCb & 6077\\
\hline
\end{tabular}
\caption{The distance between two experiments at the LHC.}
\label{table1}
\end{table}

To simplify the studies, for the detection site, we consider the ATLAS (CMS) detector as a cylinder with a diameter of 25 m (15 m). The length of the cylinder is taken to be 44 m (28.7 m). For ATLAS (CMS) detector, the outermost layer is the muon chamber, which takes O(1) fraction of the total volume of the detector, with coverage on both barrel and endcap regions.  The ATLAS (CMS) muon chamber is embedded in a magnet field of 1 (2) Tesla, orthogonal (parallel) to the beam pipe. For the LHCb experiment, the detector is asymmetric. The LHCb muon chamber can also be approximately treated as a cylinder with diameter 4.5 m and height 4 m. Lastly, ALICE does not have a dedicated muon chamber and we only consider it as a production site. In ATLAS, CMS and LHCb experiments, muon chambers are the largest components compared to other ones, such as inner tracker, ECAL and HCAL. They consequently give the largest acceptance to look for LLPs. In addition, the background in muon chambers is expected to be much cleaner. Thus in this study, we only consider LLP detection using muon chambers at these three experiments. 

The Run-III of the LHC starts in 2021, ATLAS and CMS experiments each plan to take about 150 $fb^{-1}$ data, LHCb and ALICE plan to take about 15 $fb^{-1}$ data. The high luminosity run of the LHC  will operate between 2026-2035. The total integrated luminosity at ATLAS and CMS can reach 3 $ab^{-1}$ per experiment, while LHCb and ALICE may reach a total luminosity as 500 $fb^{-1}$ \cite{{ApollinariG:2017ojx}}.

The solid  angle of a far detector to a collision point decreases quadratically with distance and it is approximately proportional to the two-third power of a detector's volume. It turns out that the ATLAS-LHCb/ALICE pair is an ideal choice among all possible combinations listed in Tab. \ref{table1}. First, the distance between ATLAS and LHCb/ALICE is the smallest. Second, the volume of ATLAS is about 3.8 times of that of CMS. At last, as shown in Fig. \ref{angle}, HV mesons and muons in final states are more concentrated along the forward-backward direction, i.e. along the beam. All these aspects exclude CMS from the optimal choice of this search. Furthermore, although the total luminosity at ATLAS is about 10 times larger than that at LHCb and ALICE, the ATLAS muon detector is much larger. It turns out that it is more efficient to use the ATLAS detector to look for LLPs produced at the LHCb and ALICE, not the other way around. Combining all these aspects, we consider using the ATLAS muon chamber to search for LLPs produced at ALICE/LHCb collision points in this study.


\section{Event Selection and Background Estimation}

Muons from HV meson decays need to be energetic enough to propagate through the magnetic field and hit multiple layers of the muon chamber at ATLAS. We impose a lower limit, 1 GeV, on muon's transverse momentum $p_T$, defined in ATLAS coordinates. 

The dominant background comes from proton-proton collisions within ATLAS experiment itself. During data taking, beam bunch crossing happens at 40 MHz at the center of the ATLAS detector. On average, there are up to 140 proton-proton collisions per beam bunch crossing. Most of the particles from these collisions are stopped in the calorimetry system between the collision point and the muon chamber, except for a moderate number of muons, and very rare random punch-through particles passing through. 
There are three features that can be used to distinguish between muons from ATLAS interaction points and muons from LLP produced from other experiments: 

$\bullet$ First, muons from LLP decays have a preferred incidence angle. The muon may not precisely point back to the collision point of LHCb or ALICE, due to the decay of the HV meson as well as its propagation in earth. Fortunately, by requiring the muon to have large momentum when it enters the ATLAS detector, i.e. $p_T >$ 1 GeV, we select muons which are sufficiently energetic so that their directions are still approximately pointing back to LHCb or ALICE. 

\begin{figure}[htbp]
 \centering
 \includegraphics[width =.5\textwidth,height=.35\textwidth]{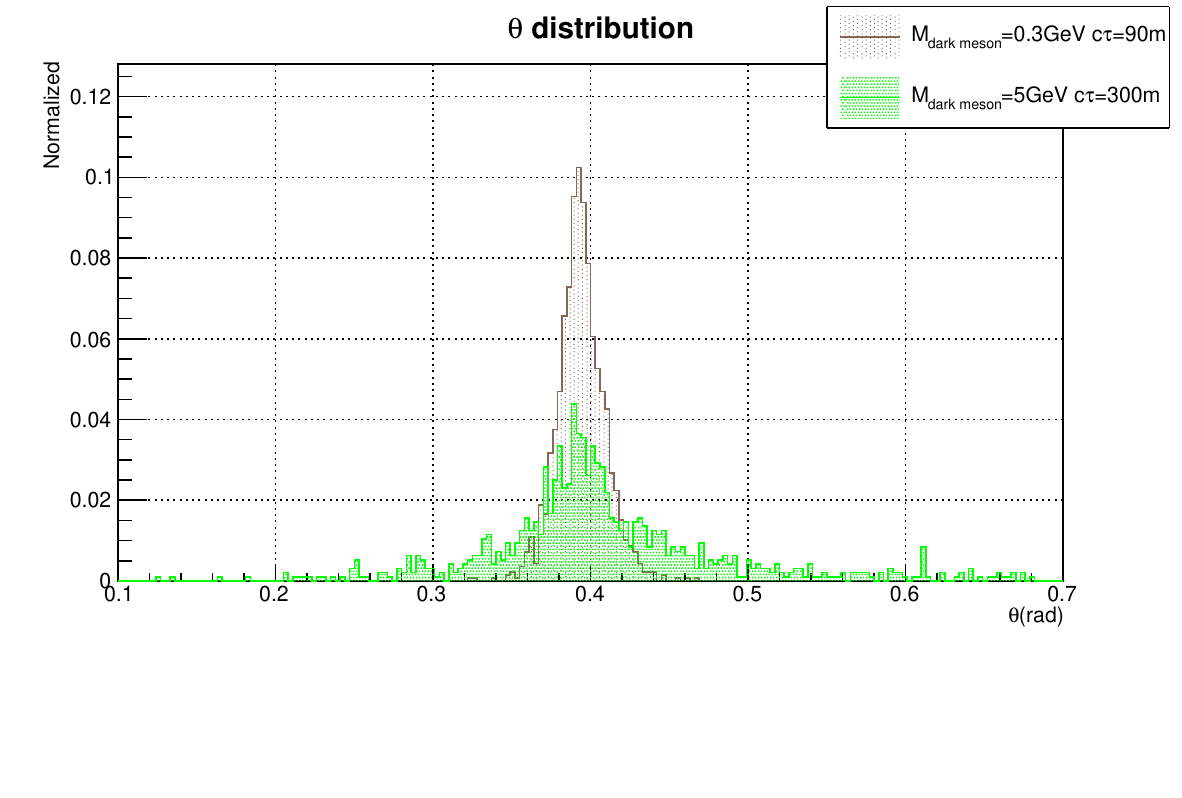}
 \caption{The $\theta$ distribution of incoming muons in the coordinate of the ATLAS detector.}
 \label{direction}
 \end{figure}

The directional information of the muon can be used to distinguish the signal from background. Taking the same set of benchmark parameters as before,  in Fig. \ref{direction}, we show the $\theta$ distribution of muons from HV meson decays in the coordinate of the ATLAS detector.  The spread of the incoming angle is $\sim 0.02$. Given the seperation between ALICE/LHCb and ATLAS as about 3300 m, the smearing on the production point is $\sim$40 m. This is similar to the size of the LHCb/ALICE experiment. Thus the muons which pass the $p_T$ threshold remain approximately pointing back to LHCb or ALICE.

$\bullet$ Second, most of muons from LLP decays have very large impact parameters, i.e. O(10) m. For muons produced at ATLAS proton-proton collision point, their impact parameters are generically smaller than O(1) mm;  

$\bullet$ Third, for a muon from LLP decay, when it passes through the detector from outside to inside, the timing sequence registered in the muon chamber is reversed to that of a muon from ATLAS proton-proton collisions.

We expect such three features of the signal can be extracted through event reconstruction, with efficiency of O(1). We assume they can be used to reduce the SM background to a negligible level.




Another concern is the background from cosmic ray muons, which can be modeled by a modified Gaisser formula~\cite{Guo:2007ug}. However, such background can be reduced to a negligible level by requiring the direction of the incoming muons to point back to the LHCb or ALICE. These experiments are about 100 meters underground and the thickness of the rock is on average of 35.6 km. The background event induced by a cosmic muon reaching a detector from these particular directions is much smaller than 1. 

There are also other possible sources of background, such as particles from environment radiations, which can be removed efficiently by the muon transverse momentum cut at 1 GeV. Thanks to the shielding of calorimetery, hit rates in muon detector is around kilo Hz/cm$^2$ at phase II~\cite{Collaboration:2285580}, with proton-proton bunch crossing rate of 40 MHz. A typical muon track registered the muon detector requires 8 good hits. For LLP search, we could in principle double this number because the muons from LLPs roughly move inward in the detector first, then run outward. With excellent position resolution of $\sim$80 $\mu$m for the muon detector, the possibility of a misconstructed track by random hits to fake our signal is negligible. 

\par
Given the facts listed above, it's a reasonable assumption that the background to our LLP signal searches is zero. 


\section{Expected Sensitivity}


In this section, we present the expected sensitivities for our benchmark model presented in Eq. (\ref{Lag}). $Z'$ mass is taken to be 200 GeV. We choose HV meson masses to be 0.3 GeV and 5 GeV to represent light and heavy hidden sectors. As discussed above, the averaged number of HV mesons, i.e. $N_{HV}$, from SH processes is model dependent, thus we choose hidden sector parameters in order to change $N_{HV}$ by a factor of 2 approximately. For $m_{HV}=0.3$ GeV, we take $N_{HV}$ to be $\sim 20$ and $\sim 40$. For $m_{HV}=5$ GeV, $N_{HV}$ is taken to be $\sim 4$ and $\sim 8$.

Since the muon direction is not always perfectly aligned with its mother HV meson direction, it does not necessarily point back to the production point. In order to accurately calculate the probability of muons in final states which can enter a far detector, we calculate the trajectory of each muon, assuming the propagation of these muons remains approximately straight in earth, as demonstrated in Fig. \ref{delphi}. 

\begin{figure}[htbp]
 \centering
 \includegraphics[width =.5\textwidth,height=.25\textwidth]{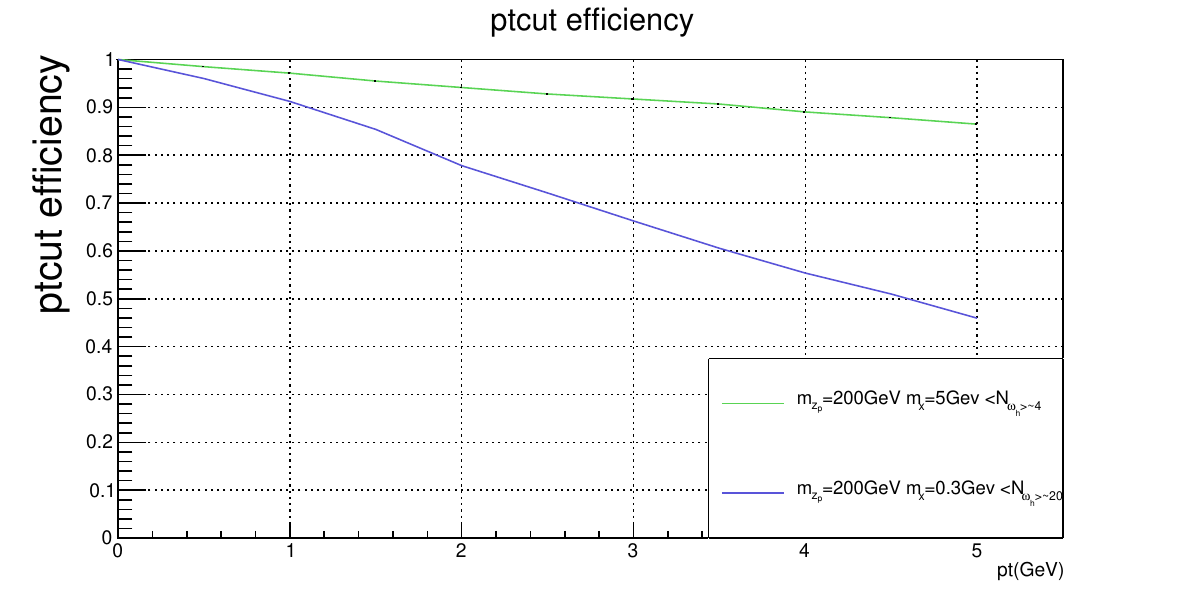}
 \caption{Event selection efficiency as a function of $p_T$ cut.}
 \label{pt}
 \end{figure}

Furthermore we impose a cut on muon $p_T$ at 1 GeV, after they penetrate the earth and reach the detector chamber, in the far detector coordinate. Such a $p_T$ cut makes sure these muons can hit multiple layers in the muon chamber. As shown in Fig. \ref{direction}, almost all muons enters the far detector at a fixed direction, i.e. $\theta\simeq 0.39$ rad. A $p_T\simeq 1$  GeV indicates a total energy as $E\simeq 2.63$ GeV. This is generally large enough to penetrate most part of the detector without being stopped, thus we expect these muon tracks can be fully reconstructed. In addition, we show how the event selection efficiency changes with the $p_T$ cut in Fig. \ref{pt}. Here we see that raising $p_T$ cut does not induce a significant change, especially for heavier HV meson mass, i.e. 5 GeV.

For muons which pass directional and kinetic cuts, the trigger efficiency is taken as $\sim$100\%. Such a choice is reasonable because a very displaced  track trigger using kalman filter at CMS has been implemented, which can trigger our signal when the beam collision is on. It has achieved an efficiency about 80\% \cite{ICHEP}.
\par

Furthermore, let us comment on a potential trigger strategy which may be used for our search. We note that the current ATLAS trigger strategy can be largely borrowed for our purpose, with a few reasonable modifications. Here we focus on the endcaps because the trigger at the barrel is much easier due to significantly lower background. 

Currently, the ATLAS muon L1 trigger is based on coincidence windows pointing to/starting from the impact point\cite{NSWTDR}.  Let us define two pointing angles: the pointing angle 1, $\Delta\theta_1$ , as the angle of the segment with respect to an ‘infinite momentum track’, i.e. a line from the interaction point of ATLAS detector to the segment’s radial position in the ATLAS New Small Wheel(NSW) detector; and pointing angle 2, $\Delta\theta_2$, as the angle of the segment with respect to an ‘infinite momentum track’, i.e. a line from the interaction point of the LLP producing detector, such as LHCb, to the segment’s radial position in the ATLAS muon Big Wheel. There is no sizable magnetic field beyond the ATLAS muon Big Wheel, thus we expect $\Delta\theta_2$ of the muons from HV meson decays is approximately zero, even for the low energy ones. Requiring muons with $\Delta\theta_2$ \textless 0.1 is a criterion to select our signal. We note that soft muons are selected even they are not energetic enough to penetrate the end-cap toroid. Meanwhile, requiring $\Delta\theta_1$ \textgreater 0.25 for each tracklet can be used to significantly remove tracks come from ATLAS collisions, which are our background. 
\par

\begin{figure}[htbp]
 \centering
 \includegraphics[width = .5\textwidth,height=.3\textwidth]{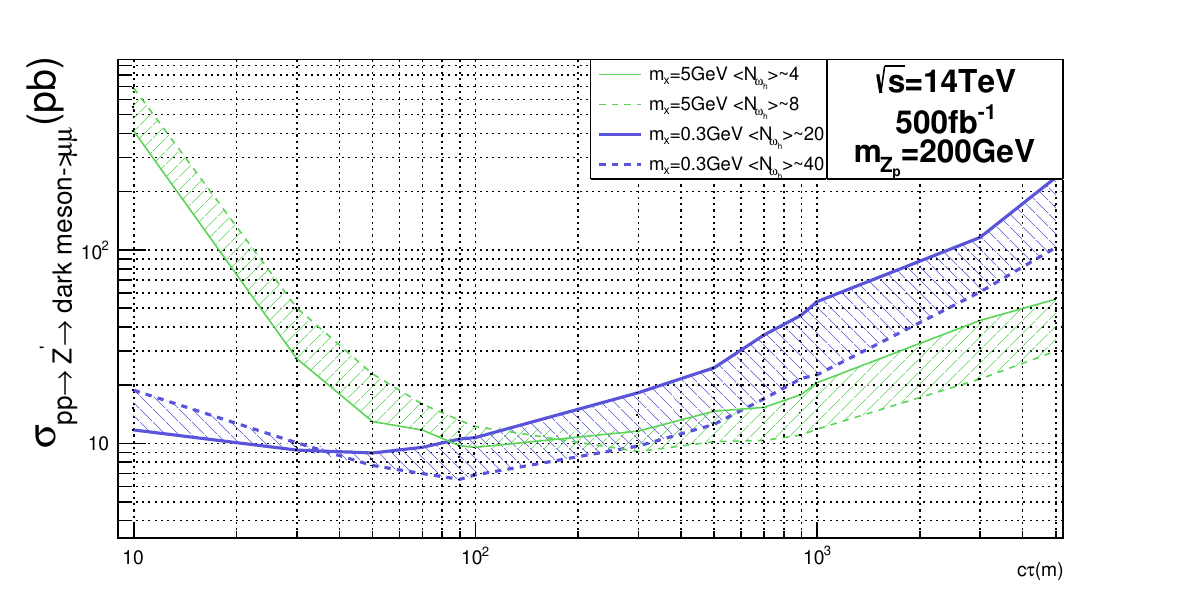}
 \caption{Expected sensitivities of our proposed search at the end of the high-luminosity run of the LHC, with different choices of benchmark parameters. LLPs are produced at LHCb or ALICE collision points. They decay to dimuon pairs and are further detected at the ATLAS detector.}
 \label{limit}
 \end{figure}

In Fig. \ref{limit}, we present the estimated sensitivities as a function of HV meson decay lifetime, for various choices of benchmark parameters. The two bands correspond to two choices of $m_{HV}$ and the boundaries of each band are for two $N_{HV}$ in each case.

Now let us briefly comment on the comparisons with other proposed search strategies. First, there are several proposals suggesting to add new modules to the LHC in order to look for LLPs, such as SHiP \cite{Alekhin:2015byh}, FASER \cite{Feng:2017uoz}, MATHUSELA \cite{Chou:2016lxi} and CODEX-b \cite{Gligorov:2017nwh}. We expect that, for our benchmark model, these newly proposed experiments can achieve a much better sensitivity compared to that of our proposed search strategy here.  However the costs of the new modules can be significant, while our proposed search only requires implementing novel trigger scheme, which is much more cost-efficient. 

Additionally, it has been proposed to use timing information at ATLAS and CMS to look for LLPs \cite{ Liu:2018wte}. This has been demonstrated to be very powerful. However, in order to generate significant time difference to distinguish signal from background, the flight direction of a daughter particle from a LLP decay needs to be quite different from that of its mother particle. This requirement is not easy to satisfy if the LLP is light, thus the study proposed in this paper is complementary to the LLP searches based on timing information.

Lastly, LHCb can also be used to search for LLPs in HV scenario \cite{Pierce:2017taw}, thanks to its unique detector configurations, especially the VELO and RICH. This search also relies on the new trigger on soft muons to be implemented at the LHCb, and it provides comparable sensitivity at large $c\tau$ as that achieved in the study presented here. As in many measurements, operating ATLAS and CMS simultaneously provides independent checks to any potential signals. Combining the results from these two search strategies, the one studied here and the one in \cite{Pierce:2017taw}, can provide powerful cross checks and further improve the sensitivity of probing the HV scenario.


\section{Order-Of-Magnitude Estimation}

To provide a cross check about our estimated sensitivity, we perform a quick order-of-magnitude estimation in this section. We consider a benchmark where the HV meson mass is 5 GeV and the averaged HV meson number in the final states is $\langle N_{\omega_h}\rangle=8$. We also set the HV meson $c\tau$ to be 300 meters, with which the best sensitivity is achieved.

The survival probability after event selections can be calculated as the product of the following factors,
\begin{eqnarray}
P_{surv} = P_{direction}P_{energy}P_{location}.
\end{eqnarray}
Here  $P_{direction}$ is the probability for an event to have a meson flying toward the correct direction, i.e. pointing to the far detector. $P_{energy}$ is the probability of a muon decayed from a HV meson who is energetic enough to penetrate a sizable depth of earth and run into the far detector. At last, $P_{location}$ is the probability for a HV meson to decay in the neighborhood of the far detector so that the daughter muons can get into the far detector.

$P_{direction}$ can be approximately calculated using the far detector solid angle. We note that using solid angle is not precise because the HV mesons are not produced isotropically. They are more focused along the forward direction. If we take the far detector size as $\sim$50 m, the solid angle is $\sim 1.9\times 10^{-5}$. Reading from our simulation, we have $P_{direction}= 10^{-5}$, which is within the order-of-magnitude estimation.

\begin{figure}[htbp]
 \centering
 \includegraphics[width =0.5\textwidth]{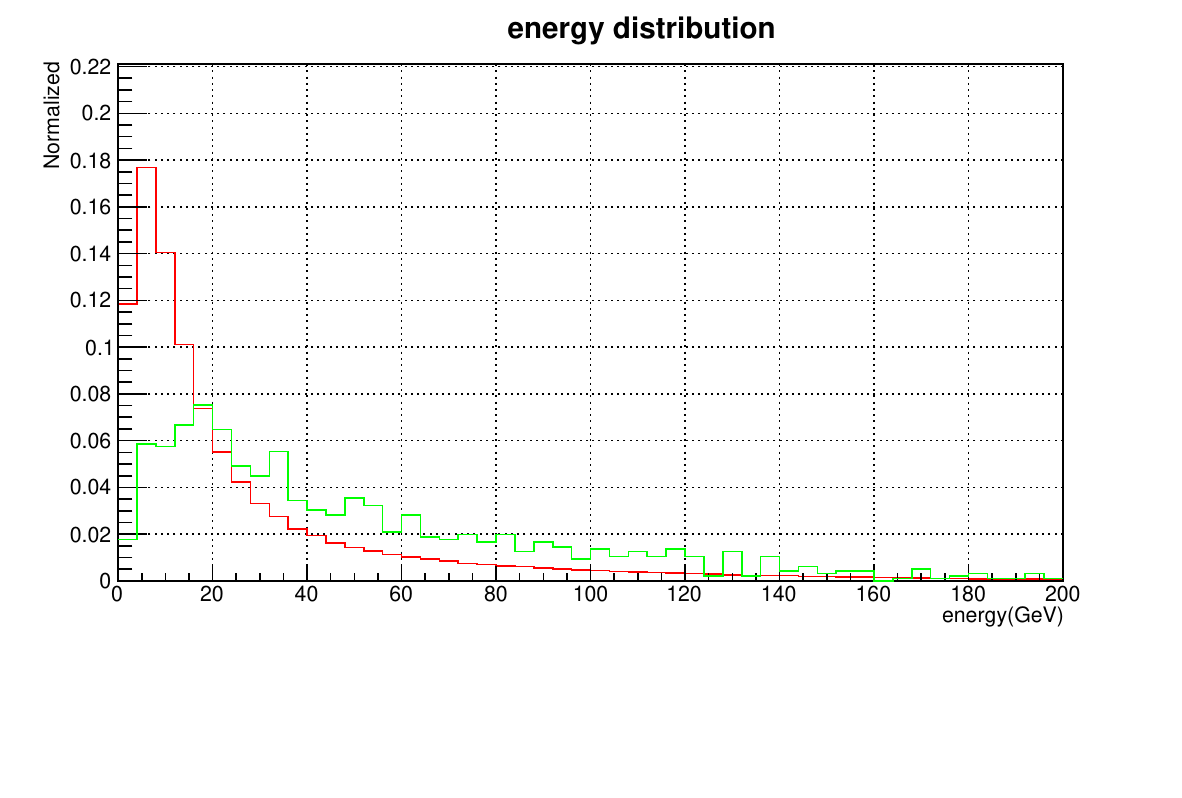}
\caption{Muon energy distribution. Red line is for muons which fly toward the far detector, and the green line is for muons which enter the far detector.}
 \label{energy2}
 \end{figure}

It is not easy to estimate $P_{energy}$ analytically, and we have to rely on the simulation. In Fig. \ref{energy2}, we show the energy distribution of the muons which fly toward the far detector versus that of muons which enter the far detector, with normalization to 1.  
 
Here we see that for all muons which fly toward the far detector, the muon energy peaks around 10 GeV. This is consistent with the generic expectation that, with good approximation, the HV mesons equally share the totally energy of the mother particle, i.e. 200 GeV in our benchmark.  However for the muons which are selected in our analysis, their energy distribution peaks at $\sim$20 GeV. Among the muons which fly toward the far detector, the probability to have a muon whose energy is larger than 20 GeV, i.e. $P_{energy}$, is about 0.39. 

The energy loss rate for a muon is about $2 \rm{MeV\times cm^2/g}$. Take the mass density of earth as 2.5 $\rm{g/cm^3}$, we find that the distance that a 20 GeV muon can penetrate in order to reach the far detector is 80 m, which is also consistent with our numerical simulation. 

Now let us calculate $P_{location}$. This can be estimated as
\begin{eqnarray}
P_{location} = e^{-3288 m/ \gamma c\tau} \frac{L}{\gamma c\tau},
\end{eqnarray}
Here $\gamma$ is the boost factor of the HV meson which run into the far detector. As discussed above, these HV mesons produce muons with typical energy 20 GeV, thus we have $\gamma\sim 8$.  $L$ is the distance to penetrate in order to reach the far detector, i.e. $L\simeq 80$ m as calculated before, and $c\tau$ is taken to be 300 m in our benchmark. Thus we have $P_{location}\simeq 0.01$.

Combine all these three factors, we have $P_{surv}\simeq 10^{-7}$. Take the total luminosity as 500 fb$^{-1}$, we need the cross section to be about 20 pb in order to get O(0.1-1) event, which is consistent with what we found in Fig. \ref{limit} within an order of magnitude.


\section{Conclusion}



In this study, we consider a novel strategy to look for LLPs. We assume LLPs are produced at one experiment, decay after flying a long distance, and their decay products are detected by the detector of another experiment. Among all possible combinations of experiments, we find that the optimal choice is to produce LLPs at the collision points of the LHCb and ALICE experiments, and the ATLAS muon chamber is used to detect muons from LLP decays. We choose a benchmark model in the HV scenario and assume a simple $Z'$-boson as the portal connecting our SM sector and the HV sector. We show that our proposed search strategy can probe interesting parameter space. It is complementary to other LLP searches, and it is cost-efficient since no new detectors are needed.

\begin{acknowledgments}
  {The authors thank Charlie Young (SLAC), Shih-Chieh Hsu (U. Washington), Yuhsin Tsai (U. of Maryland) and Lian-Tao Wang (U. Chicago) for useful discussions. We particularly thank Bart Fornal for valuable comments on the manuscript. C.Y. and H.Z. are supported by the Beijing Municipal Science \& Technology Commission, project No. Z181100004218003, Z191100007219010; National Natural Science Foundation of China, project No. 11575199, 11661141007, and Ministry of Science and Technology of China, project No. 2018YFA0403901.
  Y.Z. is supported by U.S. Department of Energy under Award Number DESC0009959.}
\end{acknowledgments}


\end{document}